\documentclass[english,notitlepage]{article}
\usepackage[T1]{fontenc}
\usepackage[latin9]{inputenc}
\usepackage{geometry}
\usepackage{xcolor}
\usepackage{pdfcolmk}
\usepackage{units}
\usepackage{url}
\usepackage{amsmath}
\usepackage{graphicx}
\usepackage[all]{xy}
\PassOptionsToPackage{normalem}{ulem}
\usepackage{ulem}

\makeatletter

\newcommand{\lyxdot}{.}

\providecolor{lyxadded}{rgb}{0,0,1}
\providecolor{lyxdeleted}{rgb}{1,0,0}

\DeclareRobustCommand{\lyxsout}[1]{\ifx\\#1\else\sout{#1}\fi}

\@ifundefined{date}{}{\date{}}
\usepackage{amsthm}
\usepackage{multicol}
\usepackage{array,multirow,graphicx}
\usepackage{adjustbox}
\usepackage{flushend}
\PassOptionsToPackage{hyphens}{url}\usepackage{hyperref}
\usepackage[left]{lineno}

\hypersetup{%
    pdfborder = {0 0 0}
}

\def\frontmatter@abstractheading{}

\renewcommand{\p@subsection}{}
\renewcommand{\p@subsubsection}{}

\usepackage{babel}

\makeatother

\usepackage{babel}
\begin{document}

\title{Snow Queen is Evil and Beautiful:\\Experimental Evidence for Probabilistic
Contextuality in Human Choices}

\author{Víctor H. Cervantes\thanks{cervantv@purdue.edu}\ \ and Ehtibar
N. Dzhafarov\thanks{to whom correspondence should be addressed (ehtibar@purdue.edu)}\\Purdue
University, USA}
\maketitle
\begin{abstract}
We present unambiguous experimental evidence for (quantum-like) probabilistic
contextuality in psychology. All previous attempts to find contextuality
in a psychological experiment were unsuccessful because of the gross
violations of marginal selectivity in behavioral data, making the
traditional mathematical tests developed in quantum mechanics inapplicable.
In our crowdsourcing experiment respondents were making two simple
choices: of one of two characters in a story (The Snow Queen by Hans
Christian Andersen), and of one of two characteristics, such as Kind
and Evil, so that the character and the characteristic chosen matched
the story line. The formal structure of the experiment imitated that
of the Einstein-Podolsky-Rosen paradigm in the Bohm-Bell version.
Marginal selectivity was violated, indicating that the two choices
were directly influencing each other, but the application of a mathematical
test developed in the Contextuality-by-Default theory, extending the
traditional quantum-mechanical test, indicated a strong presence of
contextuality proper, not reducible to direct influences. 

KEYWORDS: concept combinations, context-dependence, contextuality,
direct influences, marginal selectivity. \\\\
\end{abstract}
It is commonplace to say that human behavior is \emph{context-dependent}.
What is usually meant by this is that one's response to stimulus $S$
(performance in task $S$) depends on other stimuli (tasks) $S'$.
Asked to explain the meaning of LINE, one's answer will depend on
whether the word is preceded by CHORUS or OPENING. Visual size perception,
if interpreted as a response to retinal size, is influenced by distance
cues. In all such cases one can avoid speaking of context-dependence
by simply including the relevant elements of $S'$ into $S$: visual
size is a response to both retinal size and distance cues, the meaning
of LINE is a response to the word LINE and to the words preceding
it. J. J. Gibson's psychophysics (1950, 1960) was, essentially, a
change from understanding a percept as a response to a target stimulus
modified by context stimuli (as, e.g., in H. von Helmholtz's, 1867,
theory of unconscious inference) to a ``direct'' response to all
relevant aspects of the optical flow. 

\begin{figure}[bh]
\emph{
\[
\boxed{\xymatrix{S\ar[d]_{directly} &  & S'\ar@/_{1pc}/[dll]^{directly}\\
R
}
}
\]
}

\caption{\label{fig: direct}$R$ is a random variable interpreted as a response
to $S$: as $S$ changes, the distribution of $R$ generally changes.
It also changes as the ``context stimuli'' $S'$ change. The influences
of $S$ and $S'$ upon $R$ are both direct. }
\end{figure}
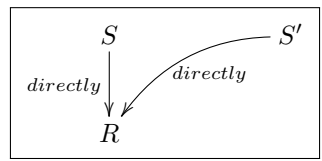

This form of context-dependence is depicted in Fig.$\,$\ref{fig: direct},
with the acknowledgement of the obvious fact that all psychological
responses are \emph{random variables}, generally varying from one
presentation to another or from one person to another (Thurstonian
cases I and II, respectively). Figure$\,$\ref{fig: direct} therefore
presents a \emph{probabilistic response} $R$ to $S$, such that its
distribution is influenced not only by $S$ but also by $S'$. This
means, of course, that the \emph{identity} of the response $R$ as
a random variable is different for different $S'$, at a fixed $S$:
one and the same random variable cannot have two different distributions.

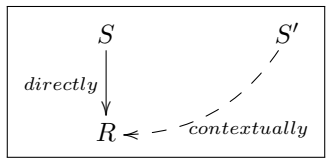
\begin{figure}
\emph{
\[
\boxed{\xymatrix{S\ar[d]_{directly} &  & S'\\
R\ar@{<--}@/_{1pc}/[urr]_{contextually}
}
}
\]
}

\caption{\label{fig: contextual}The situation when $R$, interpreted as a
response to $S$, changes its identity but not its distribution as
``context stimuli'' $S'$ change. This can be revealed by looking
at how $R$ is co-distributed with other random variables as $S'$
changes. The influence of $S$ on $R$ is direct, while influence
of $S'$ on $R$ is ``purely contextual.''}
\end{figure}

One might think that all context-dependence is of this nature: we
simply have some ``secondary'' factors influencing the distribution
of one's response to a ``primary'' one. Quantum mechanics, however,
provides striking examples of another form of context-dependence,
when the distribution of $R$ at a fixed $S$ does not change with
$S',$ but $R$ nevertheless is not one and the same random variable
at different values of $S'$. This type of context-dependence is schematically
depicted in Fig.$\,$\ref{fig: contextual}, and can be called ``purely
contextual.'' To make sure there is no logical problem here, different
random variables $R'$ and $R''$ may very well have the same distribution
(as in the case of two different fair coins). One can distinguish
them if, e.g., $R'$ is positively correlated with some random variable
$A$, $R''$ is negatively correlated with some $B$, and $A$ always
equals $B$. Obviously then, $R'$ and $R''$ cannot be one and the
same random variable, even if identically distributed. 

\begin{figure}
\begin{centering}
\includegraphics[scale=0.4]{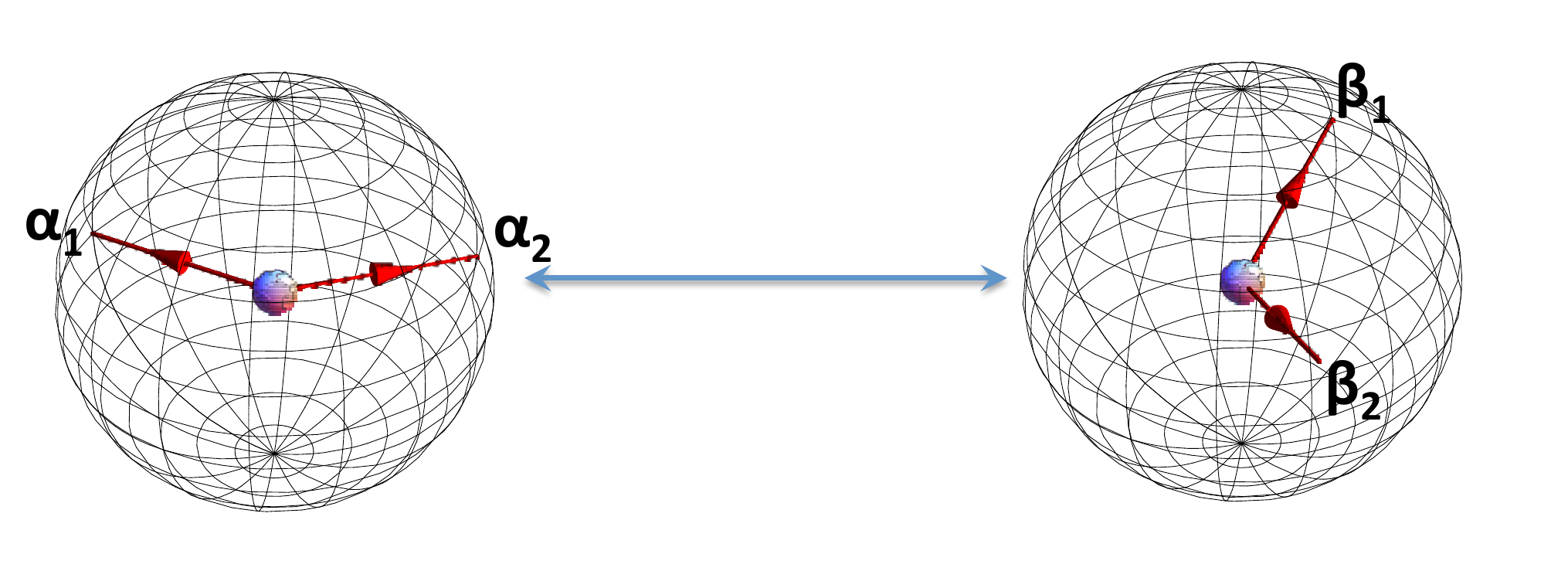}
\par\end{centering}
\caption{\label{fig:EPR/B}Einstein-Podolsky-Rosen paradigm adapted to spins
by Bohm (Bohm \& Aharonov, 1957) and famously investigated by Bell
(1964). Two spin-$\nicefrac{1}{2}$ particles (e.g., electrons) are
created in what is called a ``singlet state'' and move away from
each other. Alice measures the spin of the left particle along one
of the two axes denoted $\alpha_{1}$ and $\alpha_{2}$, Bob simultaneously
does the same for the right particle along one of the two axes denoted
$\beta_{1}$ and $\beta_{2}$. Spins are binary random variables,
with values $+1$ or $-1$. Adapted from Dzhafarov \& Kujala (2016a).}
\end{figure}

An example of pure contextuality in quantum mechanics that is especially
relevant for us (because our behavioral experiment follows its formal
structure) is the Einstein-Podolsky-Rosen paradigm in the Bohm-Bell
version (EPR/BB, Fig.$\,$\ref{fig:EPR/B}). Let us denote the spins
along axes $\alpha_{i}$ and $\beta_{j}$ by, respectively, $A_{i}$
and $B_{j}$. As it turns out, the axes can be chosen so that it is
impossible for the identity of $A_{i}$ not to depend on the choice
of $\beta_{j}$ and for the identity of $B_{j}$ not to depend on
the choice of $\alpha_{i}$. This is established by the following
reasoning. If we assume that $A_{i}$ is one and the same random variable
under $\beta_{1}$ and $\beta_{2}$ (and analogously for $B_{j}$
under $\alpha_{1}$ and $\alpha_{2}$), then we should have four jointly
distributed random variables $A_{1},A_{2},B_{1},B_{2}$, and the observed
pairs of measurements by Alice and Bob then should be derivable from
this distribution as its marginals $\left(A_{1},B_{1}\right)$, $\left(A_{1},B_{2}\right)$,
$\left(A_{2},B_{1}\right)$, and $\left(A_{2},B_{2}\right)$. If so,
these pairwise joint distributions should satisfy the following inequality,
abbreviated CHSH (Clauser, Horne, Shimony, \& Holt, 1969; Fine, 1982):
\begin{equation}
\max_{k,l\in\left\{ 1,2\right\} }\left|\sum_{i,j\in\left\{ 1,2\right\} }\mathsf{E}\left[A_{i}B_{j}\right]-2\mathsf{E}\left[A_{k}B_{l}\right]\right|-2\leq0,\label{eq: CHSH}
\end{equation}
where $\mathsf{E}$ is expected value. Now, the expected values in
the CHSH inequality can be computed for any axes $\alpha_{1},\alpha_{2},\beta_{1},\beta_{2}$
by using the principles of quantum mechanics, and it turns out that
for certain choices of these axes these expected values violate the
inequality. By reductio ad absurdum, therefore, we have to reject
the initial assumption that $A_{i}$ is the same for both choices
of $\beta_{j}$ and $B_{j}$ is the same for both choices of $\alpha_{i}$. 

In other words, $A_{i}$ and $B_{j}$ measured together are in fact
$A_{i}^{j}$ and $B_{j}^{i}$, so that, e.g., $A_{1}^{1}$ (Alice's
measurement along axis $\alpha_{1}$ when Bob has chosen axis $\beta_{1}$)
is different from $A_{1}^{2}$ (Alice's measurement along the same
axis when Bob has chosen $\beta_{2}$). However, we do not have the
same situation as in Fig.$\,$\ref{fig: direct}: Bob's choice of
an axis cannot directly influence Alice's measurement because this
choice and the measurement are simultaneous (in some inertial frame
of reference). They cannot be causally related.\footnote{\label{fn: rvs not physical}Note that the ``pure contextuality''
we have here is not a characteristic of the physical system comprised
of the two particles in Fig.$\,$\ref{fig:EPR/B}. Rather it is a
characteristic of the system of random variables representing a particular
choice of two axes by Alice and two axes by Bob. For the same two
entangled particles but a different choice of the four axes, the system
of random variables representing them may very well exhibit no contextuality.} In the past, this situation was often presented as paradoxical, with
Einstein famously referring to it as ``a spooky action at a distance.''
In fact, contextual influences involve no ``actions'' (i.e., no
transfer of energy or information). They simply reflect a fundamental
fact of probability theory, that part of the identity of a random
variable is what other random variables it is jointly distributed
with (see Dzhafarov \& Kujala, 2014a, 2016a, 2017b, for probabilistic
foundations of contextuality). A simple analogy would be the property
of being or not being ``the brightest star in the sky'' considered
part of each star's identity: the identity of a given star then can
change depending on the brightness of stars that do not influence
it directly. It is a basic but fascinating aspect of reality, fundamentally
different from direct influences in being non-causal (see Dzhafarov
\& Kujala, 2016a, for a detailed discussion).\footnote{A formal definition of a random variable in probability theory is
that it is a measurable function mapping one probability space into
another, and it is jointly distributed with any other measurable function
defined on the same domain probability space. Conversely, the set
of all random variables with which it is jointly distributed define
the domain space of this random variable, which obviously is part
of this variable's identity.}

With contextuality (or lack thereof) understood as a property of a
system of random variables describing an aspect of a physical system
(see Footnote \ref{fn: rvs not physical}), there are no known principles,
in physics or elsewhere, that would confine all contextual systems
to quantum mechanics. Quantum mechanical computations may establish
certain properties of a set of particles, and then by means of classical
probability theory one may establish that a certain system of random
variables describing these properties forms a contextual system. No
quantum mechanical computation, however, is based on contextuality
as a physical property. It is not surprising therefore that numerous
attempts were made to reveal probabilistic contextuality analogous
to the EPR/BB one outside quantum physics, in particular, in human
cognition and decision making (Aerts, 2014; Aerts et al., 2017 ; Aerts,
Gabora, \& Sozzo, 2013; Asano, Hashimoto, Khrennikov, Ohya, \& Tanaka,
2014; Bruza, Kitto, Nelson, \& McEvoy, 2009; Bruza, Kitto, Ramm, \&
Sitbon, 2015; Bruza, Wang, \& Busemeyer, 2015). The idea of constructing
a behavioral analogue of a quantum-mechanical experiment is simple:
each experimental setting (e.g., an axis chosen by Alice) is replaced
with a task of responding to a stimulus or question, and the measurement
outcome (e.g., the spin along this axis) is replaced with a response
given to this stimulus or question. With these correspondences, the
design of a behavioral experiment can be made formally identical to
that of the quantum one. For instance, in the experiment described
in Aerts, Gabora, and Sozzo (2013), the axis $\alpha_{i}$ corresponded
to the task of choosing between two animals (one pair for $i=1,$
another for $i=2$), and $\beta_{j}$ corresponded to the task of
choosing between two animal sounds (again, different pairs for $j=1$
and $j=2$). The respondent was asked to choose an animal in response
to $\alpha_{i}$ and to choose the best matching animal sound in response
to $\beta_{j}$. The expectation in this experiment was that the responses
to $\alpha_{i}$ and $\beta_{j}$ could be treated as random variables
$A_{i}$ and $B_{j}$, respectively, and the CHSH inequality (\ref{eq: CHSH})
could then be used to reveal the presence or absence of contextuality. 

Here, however, the study in question, as well as all other studies
mentioned above, faced a serious difficulty (Dzhafarov \& Kujala 2014b;
Dzhafarov, Kujala, Cervantes, Zhang, \& Jones, 2016; Dzhafarov, Zhang,
\& Kujala, 2015). The CHSH inequality (\ref{eq: CHSH}) and other
traditional contextuality tests in quantum mechanics are derived under
the assumption of ``no-signaling'' (Abramsky \& Brandenburger, 2011;
Adenier \& Khrennikov, 2017) or ``marginal selectivity'' (Dzhafarov
\& Kujala, 2014b), which is the condition ensuring that the context
does not influence random variables directly. Thus, in the classical
version of EPR/BB, the distribution of $A_{i}$ does not depend on
whether it is measured together with $B_{1}$ or $B_{2}$. Without
this condition the expression in (\ref{eq: CHSH}) would be hopelessly
confused, as the symbols it contains for random variables then would
change their meaning within the expression. In human behavior, however,
this condition is almost never satisfied: a response to a stimulus
$S$ is typically directly influenced by any stimulus $S'$ in the
temporal-spatial vicinity of $S$. For instance, in Aerts, Gabora,
and Sozzo (2013), when choosing between Tiger and Cat (task $\alpha_{2}$),
Tiger was chosen with probability 0.86 when combined with the choice
between Growls and Winnies (task $\beta_{1}$), but Tiger was only
chosen with probability 0.23 when combined with the choice between
Snorts and Meows ($\beta_{2}$). There is no way therefore one can
denote the response to $\alpha_{2}$ by $A_{2}$ and use Inequality
\ref{eq: CHSH}. The change in the distribution of the response to
$\alpha_{2}$ indicates that it is directly influenced by the choice
of the sound, while the CHSH inequality expressly excludes this possibility
(Dzhafarov \& Kujala 2014b).

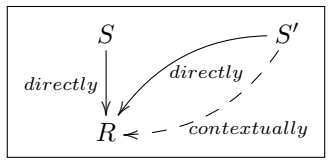
\begin{figure}
\emph{
\[
\boxed{\xymatrix{S\ar[d]_{directly} &  & S'\ar@/_{1pc}/[dll]^{directly}\\
R\ar@{<--}@/_{1pc}/[urr]_{contextually}
}
}
\]
}

\caption{\label{fig: dir+context}A combination of the situations depicted
in Figs. \ref{fig: direct} and \ref{fig: contextual}. As $S'$ changes,
the distribution of $R$ changes (i.e., $S'$ directly influences
$R$), but the identity of $R$ (revealed by looking at how $R$ is
co-distributed with other random variables) changes more than the
change in its distribution can explain. Here, influence of $S'$ on
$R$ is in part direct and in part contextual.}
\end{figure}

However, the presence of direct influences from $S'$ to $R$ does
not automatically exclude the presence of pure contextuality: it is
possible, as schematically shown in Fig.$\,$\ref{fig: dir+context},
that contextual influences coexist with direct ones. The situation
depicted in Fig.$\,$\ref{fig: contextual} is merely a special case,
when the change in the distribution of $R$ with $S'$ is nil, so
whatever change in the identity of $R$ is observed in response to
changes in $S'$, it is purely contextual. More generally, however,
one can consider the possibility that the distribution of $R$ does
change with $S'$, but the extent of this change is not sufficient
to account for the extent of the changes in $R$'s identity, as revealed
by its joint distribution with other random variables. This combined
form of context-dependence has been studied in the mathematical theory
called Contextuality-by-Default (CbD, Dzhafarov, Cervantes, \& Kujala,
2017; Dzhafarov \& Kujala, 2014a, 2016a, 2016b, 2017a; Dzhafarov,
Kujala, \& Cervantes, 2016; Dzhafarov, Kujala, \& Larsson, 2015; Kujala,
Dzhafarov, \& Larsson, 2015). 

When applied to the EPR/BB system, the logic of CbD is as follows.
One determines the maximal probability with which $A_{1}^{1}$ could
equal $A_{1}^{2}$ if the two were jointly distributed. This probability
is a measure of difference between the two distributions (the smaller
the probability the larger the difference). Analogously one determines
the maximal probabilities of $A_{2}^{1}=A_{2}^{2}$, $B_{1}^{1}=B_{1}^{2}$,
and $B_{2}^{1}=B_{2}^{2}$. If this measure of difference between
the distributions is sufficient to account for the entire difference
between the random variables $A_{1}^{1}$ and $A_{1}^{2}$, $B_{1}^{1}$
and $B_{1}^{2}$, etc., then these maximal probabilities should be
compatible with the observed joint distributions of $\left(A_{1}^{1},B_{1}^{1}\right)$,
$\left(A_{1}^{2},B_{2}^{1}\right)$, $\left(A_{2}^{1},B_{1}^{2}\right)$,
and $\left(A_{2}^{2},B_{2}^{2}\right)$. If they are, the system in
noncontextual. If they are not, then $A_{1}^{1}$ and $A_{1}^{2}$,
or $B_{1}^{1}$ and $B_{1}^{2}$, etc., have to be more dissimilar
as random variables than they are due to the difference between their
distributions. Such a system exhibits contextuality proper (``on
top of'' direct influences).\footnote{To avoid technicalities, the formulation given is far from being general
and is less than rigorous. A rigorous formulation for the EPR/BB system
involves considering \emph{maximal couplings} of $\left(A_{1}^{1},B_{1}^{1}\right)$,
$\left(A_{1}^{2},B_{2}^{1}\right)$, $\left(A_{2}^{1},B_{1}^{2}\right)$,
and $\left(A_{2}^{2},B_{2}^{2}\right)$. More complex systems require
\emph{dichotomizations} of the random variables and \emph{multimaximal
couplings} (Dzhafarov, Cervantes, \& Kujala, 2017; Dzhafarov \& Kujala,
2017a, 2017b).}

It is proved (Dzhafarov, Kujala, \& Larsson, 2015; Kujala \& Dzhafarov,
2016) that the EPR/BB system is noncontextual if and only if

\begin{equation}
\begin{array}{r}
\max_{k,l\in\left\{ 1,2\right\} }\left|\sum_{i,j\in\left\{ 1,2\right\} }\mathsf{E}\left[A_{i}^{j}B_{j}^{i}\right]-2\mathsf{E}\left[A_{k}^{l}B_{l}^{k}\right]\right|\\
\\
-\sum_{i\in\left\{ 1,2\right\} }\left|\mathsf{E}\left[A_{i}^{1}\right]-\mathsf{E}\left[A_{i}^{2}\right]\right|\\
\\
-\sum_{j\in\left\{ 1,2\right\} }\left|\mathsf{E}\left[B_{j}^{1}\right]-\mathsf{E}\left[B_{j}^{2}\right]\right|-2\leq0.
\end{array}\label{eq:modCHSH}
\end{equation}
The formula generalizes the CHSH inequality (\ref{eq: CHSH}), which
obtains if the second and third sums in the expression are zero (no-signaling
or marginal selectivity condition). When this formula was applied
to behavioral experiments imitating the EPR/BB design, all available
data (Aerts, 2014; Aerts et al., 2017 ; Aerts, Gabora, \& Sozzo, 2013;
Bruza, Kitto, Ramm, \& Sitbon, 2015; Cervantes \& Dzhafarov, 2017a;
Zhang \& Dzhafarov, 2017) were in compliance with lack of contextuality.
The same conclusion (lack of contextuality) was reached regarding
behavioral experiments with other designs (Asano, Hashimoto, Khrennikov,
Ohya, \& Tanaka, 2014; Cervantes \& Dzhafarov, 2017b; Wang \& Busemeyer,
2013; Wang, Solloway, Shiffrin, \& Busemeyer, 2014). This series of
negative results led Dzhafarov, Zhang, and Kujala (2015) and Dzhafarov,
Kujala, Cervantes, Zhang, and Jones (2016) to hypothesize that all
context-dependence in behavioral and social data may be due to direct
influences, with no contextuality proper. 

Inspection of Inequality \ref{eq:modCHSH}, however, suggests another
possibility: perhaps the correlations between $A$ and $B$ variables
in the previous attempts imitating the formal structure of the EPR/BB
experiment were not strong enough. The maximum of the first sum in
(\ref{eq:modCHSH}) is large if the four expectations $\mathsf{E}\left[A_{i}^{j}B_{j}^{i}\right]$
are large in absolute value, and one of them has the sign opposite
to the sign of the remaining three. What if this maximum were large
enough to offset the terms reflecting violations of marginal selectivity
and to make the left-hand side of the expression positive? Here, we
report an experiment in which contextuality proper is definitely established
by achieving the desired pattern of sufficiently large correlations
between $A$ and $B$ variables. 

The design of the experiment is similar to other behavioral imitations
of the EPR/BB paradigm: the choice of an axis is replaced by a choice
between two options, the options corresponding to each $\alpha$-axis
being two characters from a story, and the options corresponding to
each $\beta$-axis being two characteristics which characters from
the story may possess. The story was The Snow Queen by Hans Christian
Andersen, and, e.g., the pair $\left(\alpha_{1},\beta_{1}\right)$
was the offer to choose between Gerda and the Troll (the result being
$A_{1}^{1}$) and also to choose between Beautiful and Unattractive
($B_{1}^{1}$), so that the two choices match the story line (in which
Gerda is Beautiful and the Troll is Unattractive). The choices are
offered to many people in a crowdsourcing experiment, and the probabilities
are estimated by the proportions of people making this or that pair
of choices. The expectation is that a respondent who understands the
story line would choose a ``correct'' combination of a character
and a characteristic (e.g., either Gerda and Beautiful, or the Troll
and Unattractive). If so, the $\max$ of the first sum in Inequality
\ref{eq:modCHSH} should equal 4 (its maximal possible value), and
the presence or absence of contextuality would depend only on the
the relative proportions of people preferring one correct choice to
another. We will see, however, that a fraction of respondents, more
than 8\%, chose ``incorrect'' options.

\section*{Method}

\subsection*{Participants}

1989 participants signed up for the study on Amazon's Mechanical Turk
(Barr, J., 2005) and indicated their agreement with a standard informed
consent page in exchange for financial compensation (\$0.10). No demographics
were required nor recorded. 1799 of the participants completed the
experiment by answering the two questions posed to them. They will
be referred to below as respondents, and their responses were used
for the analysis. The number of respondents was planned to exceed
1600, estimated to be more than sufficient for construction of 99.99\%
bootstrap confidence intervals (as explained in Results). The data
were collected in January 12-14, 2017.

\subsection*{Materials and procedure}

\begin{figure}
\begin{centering}
\includegraphics[scale=0.75]{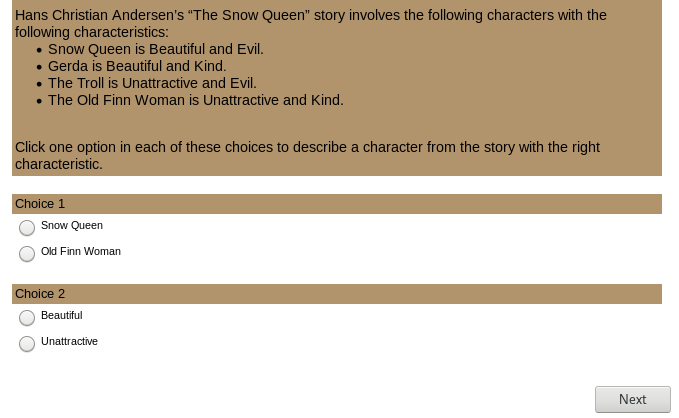}
\par\end{centering}
\caption{\label{fig: screen shot}The appearance of the computer screen for
participants assigned to Context 3.}
\end{figure}

The experiment was set up as a ``survey'' on Purdue University's
Qualtrics platform (Purdue University, 2015). Each participant was
randomly assigned to one of four conditions, referred to as contexts:
see Table \ref{tab:Contexts}. The experiment consisted in the participant
being presented with the instructions (``story line'') and, on the
same computer screen, offered to make two choices forming the context
assigned to this participant: of a character from a given pair of
characters, and of a suitable characteristic of this character from
a given pair of characteristics. For example, in Context 3 (Table
\ref{tab:Contexts}), the computer screen looked as shown in Fig.$\,$\ref{fig: screen shot},
asking to choose between Snow Queen and Old Finn Woman and to choose
between Beautiful and Unattractive, with the instruction that the
two choices had to be true to the story line (which says that Snow
Queen is Beautiful and Old Finn Woman is Unattractive).\footnote{As pointed out at the end of the Discussion section, the logic of
CbD dictates that only one context (one pair of choices) be presented
to a given respondent, dividing thereby the pool of respondents into
four groups, one responding to Context 1, another to Context 2, etc.}

\begin{table}
\centering{}\begin{adjustbox}{max width=0.9\columnwidth}
\begin{tabular}{lllc}
\hline & Character choice & Characteristic choice & N total (correct) \\ \hline 
\multirow{4}{*}{Context 1} \\
& \(\star\) Gerda & \(\star\) Beautiful & \multirow{2}{*}{447 (425)} \\
& \(\star\) Troll & \(\star\) Unattractive \\ \\
\hline 
\multirow{4}{*}{Context 2} \\
& \(\star\) Gerda & \(\star\) Kind & \multirow{2}{*}{453 (429)}  \\
& \(\star\) Troll & \(\star\) Evil \\ \\
\hline 
\multirow{4}{*}{Context 3} \\
& \(\star\) Snow Queen & \(\star\) Beautiful  & \multirow{2}{*}{446 (410)} \\
& \(\star\) Old Finn Woman & \(\star\) Unattractive \\ \\
\hline 
\multirow{4}{*}{Context 4} \\
& \(\star\) Snow Queen & \(\star\) Kind & \multirow{2}{*}{453 (388)} \\
& \(\star\) Old Finn Woman & \(\star\) Evil \\ \\
\hline
\end{tabular}
\end{adjustbox}\caption{\label{tab:Contexts}Each context consisted of two choices, between
two characters and between two characteristics. $N$ total is the
number of respondents assigned to each context (the number in parentheses
shows the subset of respondents whose answers were correct, in accordance
with the story line). }
\end{table}

\section*{Results}

We present the results first for correct responses only, and then
for all responses, with the numbers of respondents shown in Table
\ref{tab:Contexts}. In Tables \ref{tab:Correct-responses} and \ref{tab:All-responses},
we present the observed proportions for each combination of choices
in the first and second group, respectively. We refer to these tables
of proportions (or probabilities they estimate) as ``systems,''
in accordance with the terminology of ``context-content systems''
introduced in Dzhafarov \& Kujala (2016a).

\subsection*{System of correct choices}

In this system, the $\max$ of the first sum in Inequality \ref{eq:modCHSH}
equals 4 (its maximal possible value), and the presence or absence
of contextuality depends only on the the relative proportions of two
correct pairs of choices. The system is contextual on the sample level:
the left-hand side of Inequality \ref{eq:modCHSH} equals 0.452. To
evaluate how reliable this figure is, a bootstrap confidence interval
(Davison \& Hinkley, 1997) was calculated by generating $n=400,000$
resamples from each of the contexts, computing the left hand side
of Inequality \ref{eq:modCHSH} for each of them, choosing a confidence
level $C,$ and finding the $\frac{1-C}{2}$ and $1-\frac{1-C}{2}$
quantiles of their distribution. The histogram of the distribution
is shown in the upper panel of Fig.$\,$\ref{fig:histograms}. For
this system, the $99.99\%$ bootstrap confidence interval for the
left hand side of Inequality \ref{eq:modCHSH} is $[0.226,0.668]$.
The confidence needed for the bootstrap interval to cover zero exceeds
$99.999\%$ since none of the $400,000$ resamples produced a non-positive
value.

\begin{table*}
\begin{centering}
\begin{adjustbox}{max width=0.8\textwidth}
\begin{tabular}{rccccrccc}
\cline{1-4} \cline{6-9} 
&  \multicolumn{2}{c}{$B_{1}^{1}$} && &&  \multicolumn{2}{c}{$B_{2}^{1}$} \\
\multicolumn{1}{l}{Context 1} &  \multicolumn{2}{c}{\downbracefill} &&& \multicolumn{1}{l}{Context 2} &  \multicolumn{2}{c}{\downbracefill} \\
& Beautiful & Unattractive    & Mar.\ Character        & &             & Kind    & Evil    & Mar.\ Character        \\ 
\cline{1-4} \cline{6-9}  \\
\multirow{2}{*}{$A_{1}^{1} \begin{cases} \\ \end{cases}$} 
Gerda          & 0.887   & 0.000 & 0.887 & & 
\multirow{2}{*}{$A_{1}^{2} \begin{cases} \\ \end{cases}$} 
Gerda          & 0.841 & 0.000 & 0.841 \\ 
Troll          & 0.000   & 0.113 & 0.113 &  & 
Troll          & 0.000 & 0.159 & 0.159 \\ \\
\cline{1-4} \cline{6-9} \\
Mar.\ Characteristic           & 0.887   & 0.113 & & & 
Mar.\ Characteristic               & 0.841 & 0.159 & \\ \\
\cline{1-4} \cline{6-9}  
\vspace{.9em}                                     &                                &                              &                              &                       &                                     &                              &                              &                              \\ \\
\cline{1-4} \cline{6-9}  
&  \multicolumn{2}{c}{$B_{1}^{2}$} && &&  \multicolumn{2}{c}{$B_{2}^{2}$} \\
\multicolumn{1}{l}{Context 3} &  \multicolumn{2}{c}{\downbracefill} &&& \multicolumn{1}{l}{Context 4} &  \multicolumn{2}{c}{\downbracefill} \\
& Beautiful & Unattractive    & Mar.\ Character        & &             & Kind    & Evil    & Mar.\ Character        \\ 
\cline{1-4} \cline{6-9}  \\
\multirow{2}{*}{$A_{2}^{1} \begin{cases} \\ \end{cases}$} 
Snow Queen     & 0.837   & 0.000 & 0.837 &  & 
\multirow{2}{*}{$A_{2}^{2} \begin{cases} \\ \end{cases}$} 
Snow Queen     & 0.000 & 0.626 & 0.626 \\
\small{Old Finn woman} & 0.000   & 0.163 & 0.163 &  & 
\small{Old Finn woman} & 0.374 & 0.000 & 0.374 \\ \\
\cline{1-4} \cline{6-9}  \\
Mar.\ Characteristic               & 0.837   & 0.163 &  &  & 
Mar.\ Characteristic               & 0.374 & 0.627 &  \\ \\
\cline{1-4} \cline{6-9}  
\end{tabular}
\end{adjustbox}
\par\end{centering}
\caption{\label{tab:Correct-responses}Observed proportions of correct choices
for each of the four contexts. `Mar.' indicates marginal observed
proportions. To apply Inequality \ref{eq:modCHSH}, one of the two
options (no matter which) in each choice is encoded by +1, the other
by -1.}
\end{table*}

\subsection*{System of all responses}

\begin{figure}
\begin{centering}
\includegraphics[scale=0.75]{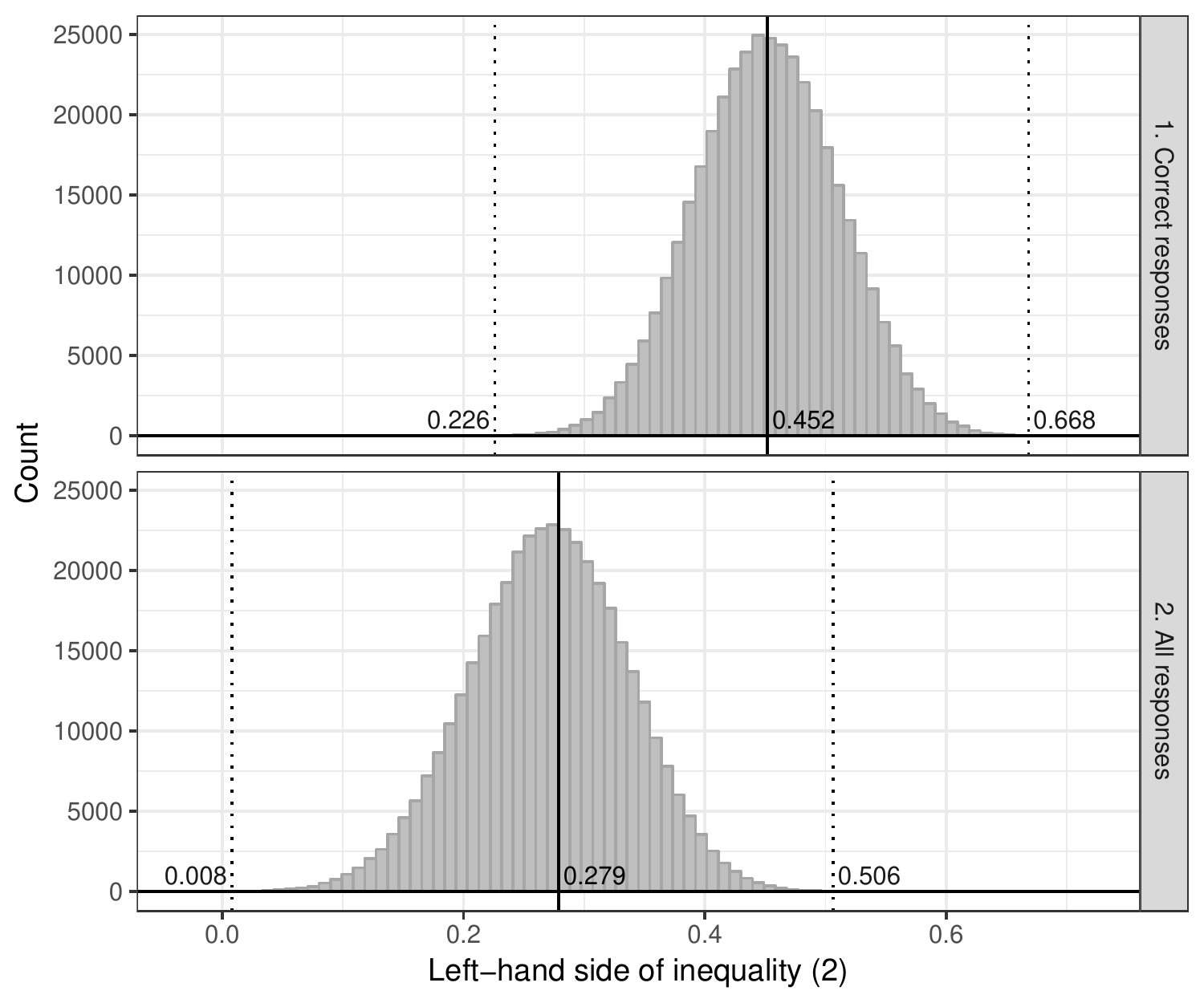}
\par\end{centering}
\caption{\label{fig:histograms}Histograms of the bootstrap values of the
left hand side of Inequality \ref{eq:modCHSH}, for correct responses
(upper panel) and for all responses (lower panel). The solid vertical
line indicates the location of the observed sample value. The vertical
dotted lines indicate the locations of the $99.99\%$ bootstrap confidence
intervals.}
\end{figure}

This system is contextual on the sample level: the left-hand side
of Inequality \ref{eq:modCHSH} equals 0.279. A bootstrap confidence
interval was calculated by generating $n=400,000$ resamples and analyzing
them in the same way as for the system of correct responses. The histogram
of the distribution of values of the left hand side of Inequality
\ref{eq:modCHSH} is shown in the lower panel of Fig.$\,$\ref{fig:histograms}.
For this system, the $99.99\%$ bootstrap confidence interval for
the left hand side of Inequality \ref{eq:modCHSH} is $[0.008,0.506]$.

\begin{table*}
\begin{centering}
\begin{adjustbox}{max width=0.8\textwidth}
\begin{tabular}{rccccrccc}
\cline{1-4} \cline{6-9} 
&  \multicolumn{2}{c}{$B_{1}^{1}$} && &&  \multicolumn{2}{c}{$B_{2}^{1}$} \\
\multicolumn{1}{l}{Context 1} &  \multicolumn{2}{c}{\downbracefill} &&& \multicolumn{1}{l}{Context 2} &  \multicolumn{2}{c}{\downbracefill} \\
& Beautiful & Unattractive    & Mar.\ Character        & &             & Kind    & Evil    & Mar.\ Character        \\ 
\cline{1-4} \cline{6-9}  \\
\multirow{2}{*}{$A_{1}^{1} \begin{cases} \\ \end{cases}$} 
Gerda          & 0.843 & 0.020 & 0.864 &  & 
\multirow{2}{*}{$A_{1}^{2} \begin{cases} \\ \end{cases}$} 
Gerda          & 0.797 & 0.035 & 0.832 \\
Troll          & 0.029   & 0.107 & 0.136 &  & 
Troll          & 0.018 & 0.150 & 0.168 \\ \\
\cline{1-4} \cline{6-9} \\
Mar.\ Characteristic           & 0.872   & 0.128 &  &  & 
Mar.\ Characteristic               & 0.815 & 0.185 & \\ \\
\cline{1-4} \cline{6-9}  
\vspace{.9em}                                     &                                &                              &                              &                       &                                     &                              &                              &                              \\ \\
\cline{1-4} \cline{6-9}  
&  \multicolumn{2}{c}{$B_{1}^{2}$} && &&  \multicolumn{2}{c}{$B_{2}^{2}$} \\
\multicolumn{1}{l}{Context 3} &  \multicolumn{2}{c}{\downbracefill} &&& \multicolumn{1}{l}{Context 4} &  \multicolumn{2}{c}{\downbracefill} \\
& Beautiful & Unattractive    & Mar.\ Character        & &             & Kind    & Evil    & Mar.\ Character        \\ 
\cline{1-4} \cline{6-9}  \\
\multirow{2}{*}{$A_{2}^{1} \begin{cases} \\ \end{cases}$} 
Snow Queen     & 0.769   & 0.011 & 0.780 & & 
\multirow{2}{*}{$A_{2}^{2} \begin{cases} \\ \end{cases}$} 
Snow Queen     & 0.135 & 0.536 & 0.671 \\
\small{Old Finn woman} & 0.070   & 0.150 & 0.220 &  & 
\small{Old Finn woman} & 0.320 & 0.009 & 0.329 \\ \\
\cline{1-4} \cline{6-9}  \\
Mar.\ Characteristic               & 0.839   & 0.161 &  &  & 
Mar.\ Characteristic               & 0.455 & 0.545 & \\ \\
\cline{1-4} \cline{6-9}  
\end{tabular}
\end{adjustbox}
\par\end{centering}
\caption{\label{tab:All-responses} Observed proportions of all choices, correct
and incorrect, for each of the four contexts. The rest is as in Table
\ref{tab:Correct-responses}.}
\end{table*}

\begin{table*}
\begin{centering}
\begin{adjustbox}{max width=0.8\textwidth}
\begin{tabular}{rccccrccc}
\cline{1-4} \cline{6-9} 
&  \multicolumn{2}{c}{$B_{1}^{1}$} && &&  \multicolumn{2}{c}{$B_{2}^{1}$} \\
\multicolumn{1}{l}{Context 1} &  \multicolumn{2}{c}{\downbracefill} &&& \multicolumn{1}{l}{Context 2} &  \multicolumn{2}{c}{\downbracefill} \\
& Beautiful & Unattractive    & Mar.\ Character        & &             & Kind    & Evil    & Mar.\ Character        \\ 
\cline{1-4} \cline{6-9}  \\
\multirow{2}{*}{$A_{1}^{1} \begin{cases} \\ \end{cases}$} 
Gerda          & 0.817   & 0.000 & 0.817 & & 
\multirow{2}{*}{$A_{1}^{2} \begin{cases} \\ \end{cases}$} 
Gerda          & 0.911 & 0.000 & 0.911 \\ 
Troll          & 0.000   & 0.183 & 0.183 &  & 
Troll          & 0.000 & 0.089 & 0.089 \\ \\
\cline{1-4} \cline{6-9} \\
Mar.\ Characteristic           & 0.817   & 0.183 & & & 
Mar.\ Characteristic               & 0.911 & 0.089 & \\ \\
\cline{1-4} \cline{6-9}  
\vspace{.9em}                                     &                                &                              &                              &                       &                                     &                              &                              &                              \\ \\
\cline{1-4} \cline{6-9}  
&  \multicolumn{2}{c}{$B_{1}^{2}$} && &&  \multicolumn{2}{c}{$B_{2}^{2}$} \\
\multicolumn{1}{l}{Context 3} &  \multicolumn{2}{c}{\downbracefill} &&& \multicolumn{1}{l}{Context 4} &  \multicolumn{2}{c}{\downbracefill} \\
& Beautiful & Unattractive    & Mar.\ Character        & &             & Kind    & Evil    & Mar.\ Character        \\ 
\cline{1-4} \cline{6-9}  \\
\multirow{2}{*}{$A_{2}^{1} \begin{cases} \\ \end{cases}$} 
Snow Queen     & 0.907   & 0.000 & 0.907 &  & 
\multirow{2}{*}{$A_{2}^{2} \begin{cases} \\ \end{cases}$} 
Snow Queen     & 0.000 & 0.696 & 0.696 \\
\small{Old Finn woman} & 0.000   & 0.093 & 0.093 &  & 
\small{Old Finn woman} & 0.304 & 0.000 & 0.304 \\ \\
\cline{1-4} \cline{6-9}  \\
Mar.\ Characteristic               & 0.907   & 0.093 &  &  & 
Mar.\ Characteristic               & 0.304 & 0.696 &  \\ \\
\cline{1-4} \cline{6-9}  
\end{tabular}
\end{adjustbox}
\par\end{centering}
\caption{\label{tab:Correct-responses-1}Hypothetical proportions of correct
choices for each of the four contexts. This system is obtained by
adding or subtracting 0.07 to/from each of the nonzero probabilities
in Table \ref{tab:Correct-responses}. }
\end{table*}

\section*{Discussion}

We have demonstrated that a contextual system of random variables
formally analogous to the EPR/BB system in quantum mechanics can be
observed in human behavior. It has been done without making the mistake
of ignoring lack of marginal selectivity in psychological data. Marginal
selectivity (or no-signaling condition), in application to the EPR/BB
system, means that the second and third sums in Inequality \ref{eq:modCHSH}
are zero. If this were the case in our experiments (e.g., if the two
correct choices of the character-characteristic pairs were made with
equal probability), the left hand side of Inequality \ref{eq:modCHSH}
for the system with correct choices would have the maximal theoretically
possible value, 2. This would make the system a so-called PR box (Popescu
\& Rohrlich, 1994), a system forbidden by laws of both classical and
quantum mechanics. There is no a priori reason why a behavioral system
could not violate boundaries established by quantum mechanics, but
the sample level contextuality value of 0.452 obtained in our experiment
for correct responses is quite moderate, well below the quantum boundary
(so-called Tsirelson bound) of $2\left(\sqrt{2}-1\right)$.\footnote{Note, however, that the derivability of the Tsirelson bound without
assuming non-signaling is not obvious and requires special investigation.} Recall that application of Inequality \ref{eq:modCHSH} and similar
formulas to all previously reported experimental data showed no contextuality
at all, leading Dzhafarov, Zhang, \& Kujala (2015) and Dzhafarov,
Kujala, Cervantes, Zhang, and Jones (2016) to consider the possibility
that all context-dependence in psychology is due to direct influences
only. This hypothesis is now falsified.\footnote{It is worth mentioning that violations of marginal selectivity or
no-signaling condition (the general CbD term being ``consistent connectedness'')
are also common in quantum physical experiments (Adenier \& Khrennikov,
2007; Khrennikov, 2017, pp. 25-28). Compared to behavioral data, however,
inconsistent connectedness in quantum mechanics is relatively small,
even when statistically significant, and with the use of CbD theory
pure contextuality can usually be established at extremely high level
of confidence (see, e.g., the analysis of experimental data in Kujala,
Dzhafarov, \& Larsson, 2015). }

Contextuality in our experiment was exhibited by both the system of
correct responses and the system of all responses, correct and incorrect.
It is not clear, however, why some respondents made incorrect choices
to begin with. The possibilities range from misunderstanding of the
instructions to deliberate non-compliance. This makes no difference
for the formal contextuality analysis, but one might consider the
legitimacy of excluding incorrect choices as outliers.

Focusing on the system of correct responses, one might wonder if a
story line that makes one of the two choices in each context rigidly
determined by the other choice (Table \ref{tab:Contexts}) may somehow
predetermine the contextuality of the system. Could the results reported
in this paper be essentially forced by the experiment's design? It
is easy to see that this is not the case. For example, in Table \ref{tab:Correct-responses-1}
all responses are correct but the system is noncontextual, with the
left-hand side of Inequality \ref{eq:modCHSH} equal to -0.004. No
superficial inspection of this system would reveal a qualitative difference
from the one in Table \ref{tab:Correct-responses}. The question should
not be therefore whether noncontextuality is compatible with the story
line, but whether the latter makes it ``rare.'' 

One way of making the meaning of ``rare'' precise is as follows.
The experimental design we use (considering only correct responses)
makes the value
\begin{equation}
\max_{k,l\in\left\{ 1,2\right\} }\left|\sum_{i,j\in\left\{ 1,2\right\} }\mathsf{E}\left[A_{i}^{j}B_{j}^{i}\right]-2\mathsf{E}\left[A_{k}^{l}B_{l}^{k}\right]\right|\label{eq: first}
\end{equation}
equal to 4, its maximal possible value. The system's (non)contextuality
therefore is determined entirely by the value of 
\begin{equation}
\sum_{i\in\left\{ 1,2\right\} }\left|\mathsf{E}\left[A_{i}^{1}\right]-\mathsf{E}\left[A_{i}^{2}\right]\right|+\sum_{j\in\left\{ 1,2\right\} }\left|\mathsf{E}\left[B_{j}^{1}\right]-\mathsf{E}\left[B_{j}^{2}\right]\right|.\label{eq: the sum}
\end{equation}
The system is contextual if and only if this expression's value is
less than 2. In the system of correct responses
\begin{equation}
\begin{array}{c}
a=\mathsf{E}\left[A_{1}^{1}\right]=\mathsf{E}\left[B_{1}^{1}\right],\\
b=\mathsf{E}\left[A_{1}^{2}\right]=\mathsf{E}\left[B_{2}^{1}\right],\\
c=\mathsf{E}\left[A_{2}^{1}\right]=\mathsf{E}\left[B_{1}^{2}\right],\\
d=\mathsf{E}\left[A_{2}^{2}\right]=-\mathsf{E}\left[B_{2}^{2}\right].
\end{array}
\end{equation}
Each of the four values $a,b,c,d$ ranges between $-1$ and $1$.
It is reasonable now to ask how probable it is that these four values
chosen ``randomly and independently'' (meaning that the quadruple
of the expected values is uniformly distributed within the 4-dimensional
cube) would yield (\ref{eq: the sum}) equal to or exceeding 2. The
answer is easily obtained by Monte Carlo simulation, and the probability
in question, i.e., the probability that a randomly created system
is noncontextual, turns out to be about 0.6667. This is hardly a ``rare''
event.

In psychological terms, the interpretation of context-dependence in
our experiment is straightforward: the meaning of such characteristics
as Kind vs Evil or Beautiful vs Unattractive is different depending
on what choice of characters is offered to ascribe these concepts
to. This difference, however, cannot be fully explained by assuming
that the impact of the character choice upon the characteristic choice
is ``direct'' (analogous to a signal propagating from Bob's measurement
to Alice's measurement). The direct influence is there without doubt,
manifested in the lack of marginal selectivity in our data, but the
context-dependence contains a component of pure contextuality. We
have no psychological terms to discern the two parts of context-dependence.
The value of contextuality analysis here is in that it provides rigorous
analytic discernments where ``ordinary'' psychological analysis
is underdeveloped or moot.

Note that the term ``direct influences'' in CbD refers to mathematical
properties of a specific system of random variables rather than to
physical or psychological mechanisms. Although the initial intuition
of direct influences involves conventional schemes with forces and
energy transfer, in the mathematical theory direct influences are
\emph{defined} by the differences between the distributions of random
variables measuring (responding to) the same property in different
contexts. In the EPR/BB system, the difference between the distributions
$A_{1}^{1}$ and $A_{1}^{2}$ is, by definition, the difference between
the direct influences exerted by $\beta_{1}$ and $\beta_{2}$ (or,
simply, the direct influence of the $\beta$) upon the measurement
of $\alpha_{1}$. If the two distributions are identical, $\beta$
exerts no direct influence, because we only think of particular random
variables and of differences in their distributions. It is perfectly
possible that the two identical distributions would differ from the
distribution of some $A_{1}^{3}$, had there been a third context
in which $\alpha_{1}$ were measured alone or together with some $\beta_{3}$.
Moreover, it is possible that a physical theory could establish that
the influences exerted by $\beta_{1}$ and $\beta_{2}$ are physically
different despite affecting the distributions of $A_{1}^{1}$ and
$A_{1}^{2}$ identically (see, e.g., Filk, 2015). Our analysis, however,
does not depend on this or that physical or psychological theory.
Even in the case of the classical EPR/BB system with two particles,
the Bohmian version of quantum mechanics allows for the possibility
of direct influences being responsible for the entire picture, albeit
defying special relativity. However, the EPR/BB system with a specific
choice of axes would remain contextual even if the Bohmian mechanics
became universally accepted. As everything else in CbD, ``direct
influence'' is not a physical term (although it may be assigned a
physical interpretation in many cases), it is a mathematical term
that is relative to the system of random variables in play.\footnote{Of course, if the systems with physically certified direct influences
that are not reflected in the differences between the distributions
were ubiquitous, the CbD analysis would be less interesting to physicists.
This is too complex an issue to discuss in a paper focusing on a single
experiment. We believe in the ``no-conspiracy'' principle reflected
in Einstein's famous ``Subtle is the Lord, but malicious He is not.''
All known to us examples of hidden direct influences are artificially
constructed on paper, with even slight modifications revealing them. }

Our experiment establishes a clear template for designing analogous
experiments aimed at pure contextuality, whether in the EPR/BB or
similar format. In the terminology of CbD, the EPR/BB system is a
cyclic system of rank 4 (Dzhafarov, Kujala, \& Larsson, 2015; Kujala,
Dzhafarov, \& Larsson, 2015). This system involves eight binary random
variables, $A_{j}^{i}$, $B_{i}^{j}$ ($i,j\in\left\{ 1,2\right\} $),
and the design maximizing the chances of this system exhibiting contextuality
(``on top of'' direct influences) is as follows. Label the values
of all the random variables $+1$ and $-1$ and create a ``story
line'' in which +1 of $A_{j}^{i}$ and +1 of $B_{i}^{j}$ are associated
with a very high probability in three out of four pairs $\left(A_{j}^{i},B_{i}^{j}\right)$,
and with a very low probability in the fourth pair (or vice versa).
For other cyclic systems (say, of ranks 3 or 5) the criteria of contextuality
are similar to (\ref{eq:modCHSH}), and the design can be constructed
similarly.

An important feature of the design is that each respondent should
be assigned to a single context only, instead of asking each of them
to make (in the case of the EPR/BB system) all four pairs of choices
$\left(\alpha_{1},\beta_{1}\right),\ldots,\left(\alpha_{2},\beta_{2}\right)$,
whether presented simultaneously, in a fixed order, or a variable
order. The reason for this is that making all four pairs of choices
would have created an empirical joint distribution of the eight random
variables in play, 
\[
A_{1}^{1},B_{1}^{1},A_{1}^{2},B_{2}^{1},A_{2}^{1},B_{1}^{2},A_{2}^{2},B_{2}^{2},
\]
contravening the logic of CbD in which different contexts are mutually
exclusive, and different pairs $\left(A_{i}^{j},B_{j}^{i}\right)$
are not jointly distributed (are \emph{stochastically unrelated} to
each other). Contextuality analysis consists in finding out whether
a joint distribution can be \emph{imposed} on these eight random variables,
subject to certain constraints (maximality of the probabilities of
$A_{1}^{1}=A_{1}^{2}$, $B_{1}^{1}=B_{1}^{2}$, etc.). For this analysis
an empirical joint distribution involving, say, $A_{1}^{1}$ and $A_{1}^{2}$
would be a nuisance relation. It would have to be ignored, and an
additional theory would be required to know how the ignored relations
affect the results of the analysis. Consider, e.g., the fact that
every given choice (e.g., between Gerda and Troll) in the EPR/BB system
enters in two different contexts. The respondent would normally remember
her previous choice when facing it the second time, albeit in combination
with another pair of characteristics (which in turn, will appear once
again, in combination with another pair of characters). It is clear
that the respondent's choice would depend on the previously made one
in some complex way (e.g., the strategy may be adopted to always repeat
it, or to always choose a new option). This would affect the marginal
distributions of the choices in some unknown way. Note that our design
is not different from how the measurements are made in the quantum-mechanical
EPR/BB system, where only one pair of measurements can be performed
on a given pair of entangled particles.

Jerome Busemeyer (personal communication, November 2017) mentioned
to us that the ``respondents'' in our design need not be people,
they can be any entities to which $A$ and $B$ properties can be
probabilistically assigned. This observation points at ways of searching
for contextual systems outside both quantum mechanics and psychology.

\paragraph*{Acknowledgments.}

This research has been supported by AFOSR grant FA9550-14-1-0318.
The authors thank Janne V. Kujala and Emmanuel Pothos for valuable
critical suggestions. 

\section*{REFERENCES}

\setlength{\parindent}{0cm}\everypar={\hangindent=15pt}
\begin{enumerate}
\item Abramsky, S., \& Brandenburger, A. (2011). The sheaf-theoretic structure
of non-locality and contextuality. New Journal of Physics, 13(11),
113036. \url{https://doi.org/10.1088/1367-2630/13/11/113036} 
\item Adenier, G. and Khrennikov, A. (2007). Is the fair sampling assumption
supported by EPR experiments?, Journal of Physics B: Atomic, Molecular,
Optical Physics, 40, 131\textendash 141. \url{https://doi.org/10.1088/0953-4075/40/1/012}
\item Adenier, G., \& Khrennikov, A. Y. (2017). Test of the no-signaling
principle in the Hensen \textquotedblleft loophole-free CHSH experiment.\textquotedblright{},
Fortschritte der Physik, 65, 1600096. \url{https://doi.org/10.1002/prop.201600096}
\item Aerts, D. (2014). Quantum theory and human perception of the macro-world.
Frontiers in Psychology, 5, 1\textendash 19. \url{https://doi.org/10.3389/fpsyg.2014.00554} 
\item Aerts, D., Arguëlles, J. A., Beltran, L., Geriente, S., de Bianchi,
M. S., Sozzo, S., \& Veloz, T. (2017). Spin and wind directions I:
Identifying entanglement in nature and cognition. Foundations of Science. \url{https://doi.org/10.1007/s10699-017-9528-9}
\item Aerts, D., Gabora, L., \& Sozzo, S. (2013). Concepts and their dynamics:
A quantum-theoretic modeling of human thought. Topics in Cognitive
Science, 5(4), 737\textendash 772. \url{https://doi.org/10.1111/tops.12042} 
\item Asano, M., Hashimoto, T., Khrennikov, A. Y., Ohya, M., \& Tanaka,
Y. (2014). Violation of contextual generalization of the Leggett-Garg
inequality for recognition of ambiguous figures. Physica Scripta,
T163, 14006. \url{https://doi.org/10.1088/0031-8949/2014/T163/014006} 
\item Barr, J. (2005). Amazon\textquoteright s Mechanical Turk: The First
Three Weeks. Retrieved from \url{https://aws.amazon.com/blogs/aws/amazons_mechani/} 
\item Bell, J. S. (1964). On the Einstein-Podolsky-Rosen paradox. Physics,
1(3), 195\textendash 200. 
\item Bohm, D., \& Aharonov, Y. (1957). Discussion of experimental proof
for the paradox of Einstein, Rosen, and Podolsky. Physical Review,
108(4), 1070\textendash 1076. \url{https://doi.org/10.1103/PhysRev.108.1070 }
\item Bruza, P. D., Kitto, K., Nelson, D., \& McEvoy, C. (2009). Is there
something quantum-like about the human mental lexicon? Journal of
Mathematical Psychology, 53(5), 362\textendash 377. \url{https://doi.org/10.1016/j.jmp.2009.04.004} 
\item Bruza, P. D., Kitto, K., Ramm, B. J., \& Sitbon, L. (2015). A probabilistic
framework for analysing the compositionality of conceptual combinations.
Journal of Mathematical Psychology, 67, 26\textendash 38. \url{https://doi.org/10.1016/j.jmp.2015.06.002} 
\item Bruza, P. D., Wang, Z., \& Busemeyer, J. R. (2015). Quantum cognition:
a new theoretical approach to psychology. Trends in Cognitive Sciences,
19(7), 383\textendash 393. \url{https://doi.org/10.1016/j.tics.2015.05.001} 
\item Cervantes, V.H., \& Dzhafarov, E.N. (2017a). Exploration of contextuality
in a psychophysical double-detection experiment. In J. A. de Barros,
B. Coecke, E. Pothos (Eds.), Quantum Interaction. LNCS (Vol. 10106,
pp. 182-193). Dordrecht: Springer. . \url{https://doi.org/10.1007/978-3-319-52289-0_15}
\item Cervantes, V.H., \& Dzhafarov, E.N. (2017b). Advanced analysis of
quantum contextuality in a psychophysical double-detection experiment.
Journal of Mathematical Psychology 79, 77-84. \url{https://doi.org/10.1016/j.jmp.2017.03.003}
\item Clauser, J. F., Horne, M. A., Shimony, A., \& Holt, R. A. (1969).
Proposed experiment to test local hidden-variable theories. Physical
Review Letters, 23, 880\textendash 884. \url{https://doi.org/10.1103/PhysRevLett.23.880} 
\item Davison, A. C., \& Hinkley, D. V. (1997). Bootstrap Methods and their
Application (1st ed.). New York, NY, USA: Cambridge University Press. 
\item Dzhafarov, E. N., Cervantes, V. H., \& Kujala, J. V. (2017). Contextuality
in canonical systems of random variables. Philosophical Transactions
of the Royal Society A, 375, 20160389. \url{https://doi.org/10.1098/rsta.2016.0389} 
\item Dzhafarov, E. N., \& Kujala, J. V. (2014a). Contextuality is about
identity of random variables. Physica Scripta, 2014(T163), 14009.
\url{https://doi.org/10.1088/0031-8949/2014/T163/014009} 
\item Dzhafarov, E. N., \& Kujala, J. V. (2014b). On selective influences,
marginal selectivity, and Bell/CHSH inequalities. Topics in Cognitive
Science, 6(1), 121\textendash 128. \url{https://doi.org/10.1111/tops.12060} 
\item Dzhafarov, E. N., \& Kujala, J. V. (2016a). Context-content systems
of random variables: The Contextuality-by-Default theory. Journal
of Mathematical Psychology, 74, 11\textendash 33. \url{https://doi.org/10.1016/j.jmp.2016.04.010} 
\item Dzhafarov, E. N., \& Kujala, J. V. (2016b). Conversations on contextuality.
In E. N. Dzhafarov, S. Jordan, R. Zhang, \& V. H. Cervantes (Eds.),
Contextuality from Quantum Physics to Psychology (pp. 1\textendash 22).
Singapore: World Scientific.
\item Dzhafarov, E. N., \& Kujala, J. V. (2017b). Probabilistic foundations
of contextuality. Fortschritte Der Physik, 65, 1600040. \url{https://doi.org/10.1002/prop.201600040} 
\item Dzhafarov, E. N., \& Kujala, J. V. (2017a). Contextuality-by-Default
2.0: Systems with Binary Random Variables. In J. A. de Barros, B.
Coecke, \& E. Pothos (Eds.), Quantum Interaction. LNCS (Vol. 10106,
pp. 16\textendash 32). Dordrecht: Springer. \url{https://doi.org/10.1007/978-3-319-52289-0_2}
\item Dzhafarov, E. N., Kujala, J. V., \& Cervantes, V. H. (2016). Contextuality-by-Default:
A brief overview of ideas, concepts, and terminology. In H. Atmanspacher,
T. Filk, \& E. Pothos (Eds.), Quantum Interaction. LNCS (Vol. 9535,
pp. 12\textendash 23). Dordrecht: Springer. \url{https://doi.org/10.1007/978-3-319-28675-4_2}
\item Dzhafarov, E. N., Kujala, J. V., Cervantes, V. H., Zhang, R., \& Jones,
M. (2016). On contextuality in behavioural data. Philosophical Transactions
of the Royal Society A, 374, 20150234. \url{https://doi.org/10.1098/rsta.2015.0234} 
\item Dzhafarov, E. N., Kujala, J. V., \& Larsson, J.-Å. (2015). Contextuality
in three types of quantum-mechanical systems. Foundations of Physics,
45(7), 762\textendash 782. \url{https://doi.org/10.1007/s10701-015-9882-9} 
\item Dzhafarov, E. N., Zhang, R., \& Kujala, J. V. (2015). Is there contextuality
in behavioural and social systems? Philosophical Transactions of the
Royal Society A, 374, 20150099. \url{https://doi.org/10.1098/rsta.2015.0099} 
\item Filk, T. 2015. It is the theory which decides what we can observe.
In E. N. Dzhafarov, S. Jordan, R. Zhang, \& V. H. Cervantes (Eds.),
\emph{Contextuality from Quantum Physics to Psychology} (pp. 77-92),
New Jersey:World Scientific Press.
\item Fine, A. (1982). Hidden variables, joint probability, and the Bell
inequalities. Physical Review Letters, 48(5), 291\textendash 295.
\url{https://doi.org/10.1103/PhysRevLett.48.291} 
\item Gibson, J. J. (1950). The perception of the visual world. Boston:
Houghton Mifflin. 
\item Gibson, J. J. (1960). The concept of the stimulus in psychology. American
Psychologist, 15(11), 694\textendash 703. \url{https://doi.org/10.1037/h0047037} 
\item Helmholtz, H. von. (1867). Handbuch der physiologischen Optik. Leipzig:
Voss. Retrieved from \url{https://ia902707.us.archive.org/12/items/handbuchderphysi00helm/handbuchderphysi00helm.pdf} 
\item Khrennikov, A. (2017). Probability and Randomness: Quantum versus
Classical. Berlin: Springer
\item Kujala, J. V., \& Dzhafarov, E. N. (2016). Proof of a conjecture on
contextuality in cyclic systems with binary variables. Foundations
of Physics, 46, 282\textendash 299. \url{https://doi.org/10.1007/s10701-015-9964-8} 
\item Kujala, J. V., Dzhafarov, E. N., \& Larsson, J-Å. (2015). Necessary
and sufficient conditions for an extended noncontextuality in a broad
class of quantum mechanical systems. Physical Review Letters, 115(15),
150401. \url{https://doi.org/10.1103/PhysRevLett.115.150401} 
\item Popescu, S., \& Rohrlich, D. (1994). Quantum nonlocality as an axiom.
Foundations of Physics, 24(3), 379\textendash 385. \url{https://doi.org/10.1007/BF02058098} 
\item Purdue University. (2015). Teaching and Learning Technologies - Qualtrics.
Retrieved from \url{https://www.itap.purdue.edu/learning/tools/qualtrics.html} 
\item Wang, Z., \& Busemeyer, J. R. (2013). A quantum question order model
supported by empirical tests of an a priori and precise prediction.
Topics in Cognitive Science, 5(4), 689\textendash 710. \url{https://doi.org/10.1111/tops.12040} 
\item Wang, Z., Solloway, T., Shiffrin, R. M., \& Busemeyer, J. R. (2014).
Context effects produced by question orders reveal quantum nature
of human judgments. Proceedings of the National Academy of Sciences,
111(26), 9431\textendash 9436. \url{https://doi.org/10.1073/pnas.1407756111} 
\item Zhang, R., \& Dzhafarov, E.N. (2017). Testing contextuality in cyclic
psychophysical systems of high ranks. In J. A. de Barros, B. Coecke,
E. Pothos (Eds.) Quantum Interaction. LNCS (Vol. 10106, pp. 151\textendash 162).
Dordrecht: Springer. \url{https://doi.org/10.1007/978-3-319-52289-0_12}
\end{enumerate}

\end{document}